\RequirePackage{fix-cm}
\documentclass{svjour3}                     % onecolumn (standard format)
\smartqed  % flush right qed marks, e.g. at end of proof
\usepackage{graphicx}
\usepackage{amssymb,amsmath}
\begin{document}

\title{Seeing opportunity in every difficulty%\thanks{Grants or other notes
}
\subtitle{Protecting information with weak value techniques}


\author{George~C.~Knee         \and
        G.~Andrew~D.~Briggs%etc.
}


\institute{G.C.~Knee \at
              Department of Physics, University of Warwick, Coventry CV4 7AL, UK \\
              Department of Materials, University of Oxford, Oxford OX1 3PH, UK\\
              Tel.: +44(0)24 76150943\\
              https://orcid.org/0000-0002-7203-0271\\
              \email{gk@physics.org}           %  \\
           \and
           G.A.D.~Briggs \at
              Department of Materials, University of Oxford, Oxford OX1 3PH, UK
}

\date{Received: 31 March 2018 / Accepted: 23 April 2018}

\maketitle

\begin{abstract}
A weak value is an effective description of the influence of a pre and postselected `principal' system on another `meter' system to which it is weakly coupled. Weak values can describe anomalously large deflections of the meter, and deflections in otherwise unperturbed variables: this motivates investigation of the potential benefits of the protocol in precision metrology. 
We present a visual interpretation of weak value experiments in phase space, enabling an evaluation of the effects of three types of detector noise as `Fisher information efficiency' functions. These functions depend on the marginal distribution of the Wigner function of the `meter', and give a unified view of the weak value protocol as a way of protecting Fisher information from detector imperfections. This approach explains why weak value techniques are more effective for avoiding detector saturation than for mitigating detector jitter or pixelation.
\end{abstract}

\section{Introduction}
Professor Izumi Tsutsui has identified three categories of research into weak values~\cite{Tsutsui2015}. The first works towards an understanding of the formal and mathematical aspects of the theory. Taking $\hbar=1$ , a derivation similar to the original one by Aharonov, Albert and Vaidman (AAV)~\cite{AharonovAlbertVaidman1988}:
\begin{eqnarray}
\langle f| e^{-ig\hat{A}\hat{k}}|i\rangle|m\rangle\approx \langle f|i\rangle e^{-igA_w\hat{k}}|m\rangle;\qquad A_w = \frac{\langle f | \hat{A} | i \rangle}{\langle f | i \rangle}
\label{wva}
\end{eqnarray}
involves the definition of a new quantity $A_w$, the weak value. It may be thought of as a generalisation of the eigenvalue of $\hat{A}$~\cite{Vaidman2017}. As can be seen from the above expression, it emerges only through a weak coupling ($g\ll1$) of a principal system, preselected into $|i\rangle$ and preselected into $|f\rangle$,  with a `meter' degree of freedom, initialised in $|m\rangle$. This generalisation leaves us with an economical, effective description of meter solely in terms of the generally complex and unbounded quantity $A_w$. The physical effect of the imaginary component~\cite{Jozsa2007}, the breakdown in the aforementioned approximations~\cite{DuckStevensonSudarshan1989,KofmanAshhabNori2012}, and the full-order, exact evolution of the meter~\cite{WuLi2011} all fall within the purview of Tsutsui's first category and have been the subject of many studies. Ref~\cite{KofmanAshhabNori2012} is a comprehensive treatment of nonperturbative analogues of Equation (\ref{wva}) featuring various coupling Hamiltonians and meter states. Insofar as research in this category is mathematical it is uncontroversial, and the discussion mainly focusses on the approximations to be made and the range of their validity.

The fruits of this first category of research feed into the other two. The second category considers the foundational aspects of weak values. They have been associated with quantum paradoxes~\cite{AharonovRohrlich2005}, with negative probabilities, and with macrosopic realism~\cite{GogginAlmeidaBarbieri2011} (as has the wider topic of pre and post-selected measurements~\cite{AharonovRohrlich2005,GeorgeRobledoMaroney2013}). Weak values have been argued to provide an operational definition of the quantum wavefunction~\cite{LundeenSutherlandPatel2011}, and to reveal particle trajectories of Bohmian mechanics~\cite{KocsisBravermanRavets2011}. The question of whether weak values are quantum mechanical has also been debated~\cite{FerrieCombes2014a}. It is perhaps unsurprising that this second category of research has enjoyed a large deal of disagreement and controversy, since it involves speculation into the meaning of weak values, and indeed into the meaning of quantum mechanics itself (where there remain open questions on the interpretation of the formalism and the peculiar role of measurement~\cite{BriggsButterfieldZeilinger2013}). 

The third category ought to be far more clear-cut, for it deals with the use of weak values and associated techniques in technology, specifically for parameter estimation. Here, one might expect, the question of whether the technique is genuinely useful or not would be a matter of fact: unforgiving and unquestionable. Why then, has this line of research also been afflicted by apparent contradictions and disagreements? The third category was ignited by an experiment in which a weak value technique was used to measure the spin-Hall effect of light~\cite{HostenKwiat2008}. The interaction between the polarisation and transverse momentum of a light beam at an interface may be modelled using Equation~(\ref{wva}), where the coupling constant $g$ is considered an unknown parameter to be investigated. Using clever pre and postselected polarization states the lateral displacements could be measured with a sensitivity of around 1 Angstrom. Other impressive experiments include the estimation of the angular tilt of a mirror in a Sagnac interferometer to a precision of a few hundred femtoradians~\cite{DixonStarlingJordan2009}. From 2013 onwards, however, theoretical treatments of the weak value technique called into doubt whether there was any true advantage over standard strategies~\cite{KneeBriggsBenjamin2013,TanakaYamamoto2013,KneeGauger2014,FerrieCombes2014}, especially from the point of view of parameter estimation. There were also a slew of theoretical~\cite{LeeTsutsui2014,AlvesEscherMatos-Filho2015} and experimental~\cite{VizaMartinez-RinconAlves2015} papers arguing for the merits of weak value techniques. A review of these results is given in Ref~\cite{KneeCombesFerrie2016}. 

A central tool in our approach is to use Fisher information, rather than the more commonly employed signal-to-noise ratio. Fisher information is a way of measuring the amount of information that an observable random variable such as $x$ or $k$ carries about an unknown parameter upon which the probability of $x$ or $k$ depends. In AAV's original work~\cite{AharonovAlbertVaidman1988}, the subject of investigation is the operator $\hat{A}$ itself. The canonical example they gave was of a spin 1/2 particle traversing a magnetic field gradient, as in the famous Stern-Gerlach experiment (where the coupling between the spin and the field gradient causes a bream of such particles to be deflected `up' or `down' depending on their spin state). The title of the paper trumpeted the astonishing result that $A_w$ could take very large values (say 100), where usually an eigenvalue bounded by $\pm 1/2$ would appear.  From the parameter estimation perspective, however, the emphasis is rather more on the coupling constant $g$: in the Stern-Gerlach setup this corresponds to the magnitude of the magnetic field gradient, and generally controls the `strength' of the measurement via the degree of correlation that is built up between principal system and meter. Since the weak value $A_w$ multiplies $g$ in Equation (\ref{wva}), it controls the information about $g$ that can be extracted in the subsequent detection of the meter. %The idea may be readily grasped by considering how the uncertainty in the measured variable would be propagated to a uncertainty in $g$ % In weak value amplification, the parameter whose value is to be measured corresponds to the operator A. This is coupled to the meter with strength g, and g is often taken as a proxy for the quantity to be measured.

The lively debate in the third category has left open the question whether and how weak value techniques can offer precision advantages. The arguments against, in a nutshell, state that the postselected meters carry an amplified amount of Fisher information per measured data point, but this is more than cancelled by the low probability of the signal for each potential data point surviving the post-selection process. So the arguments seem quite clear-cut after all -- but such a summary belies some of the subtleties in evaluating metrological merit. Firstly, there is the issue of technical noise: as long as postselection is performed in the experiment before any ultimate detection (say with a polariser), rather than merely implemented as voluntary data loss, the weak value technique has characteristics that physically distinguish it from regular approaches. This means that certain types of detector noise may favour one approach over the other, potentially reversing the conclusions reached for ideal detection: although it is then difficult to say in full generality when advantages may be had, and the situation must be judged mostly on a case by case basis. Secondly, there is the issue of resource-counting. This is an ever present question in metrology, and is informed by the `cost' that is associated with each component of an experiment. The figure of merit (e.g. the precision of the method) is often given `per unit cost', as in `Fisher information per photon'. Since the constraints on the cost depend on the nature of the experiment, and can even vary from laboratory to laboratory, any general claims will necessarily depend on the details. Nevertheless we will discuss a scenario that will end up favouring weak value techniques: namely, when system-meter pairs (e.g. photons) are more costly to detect than to create.

In this paper we present a new, visual way of intuiting the utility of weak values. The main tool we use is the Wigner function: a quasi-probability distribution over phase space. We hope our approach may help bring clarity to an otherwise confusing field. We will not provide a comprehensive treatment of all possible implementations of the weak value protocol, but instead make choices that simplify the arguments while remaining faithful to the original approach and also representing most experiments so far. In our treatment of three types of technical noise -- namely transverse jitter, pixelation and saturation -- we rely heavily on theoretical results that have been previously reported~\cite{KneeGauger2014,KofmanAshhabNori2012,HarrisBoydLundeen2017} but require some re-application in our new setting.

\section{A simple model}
\begin{figure}
\centering
\includegraphics[width=16cm]{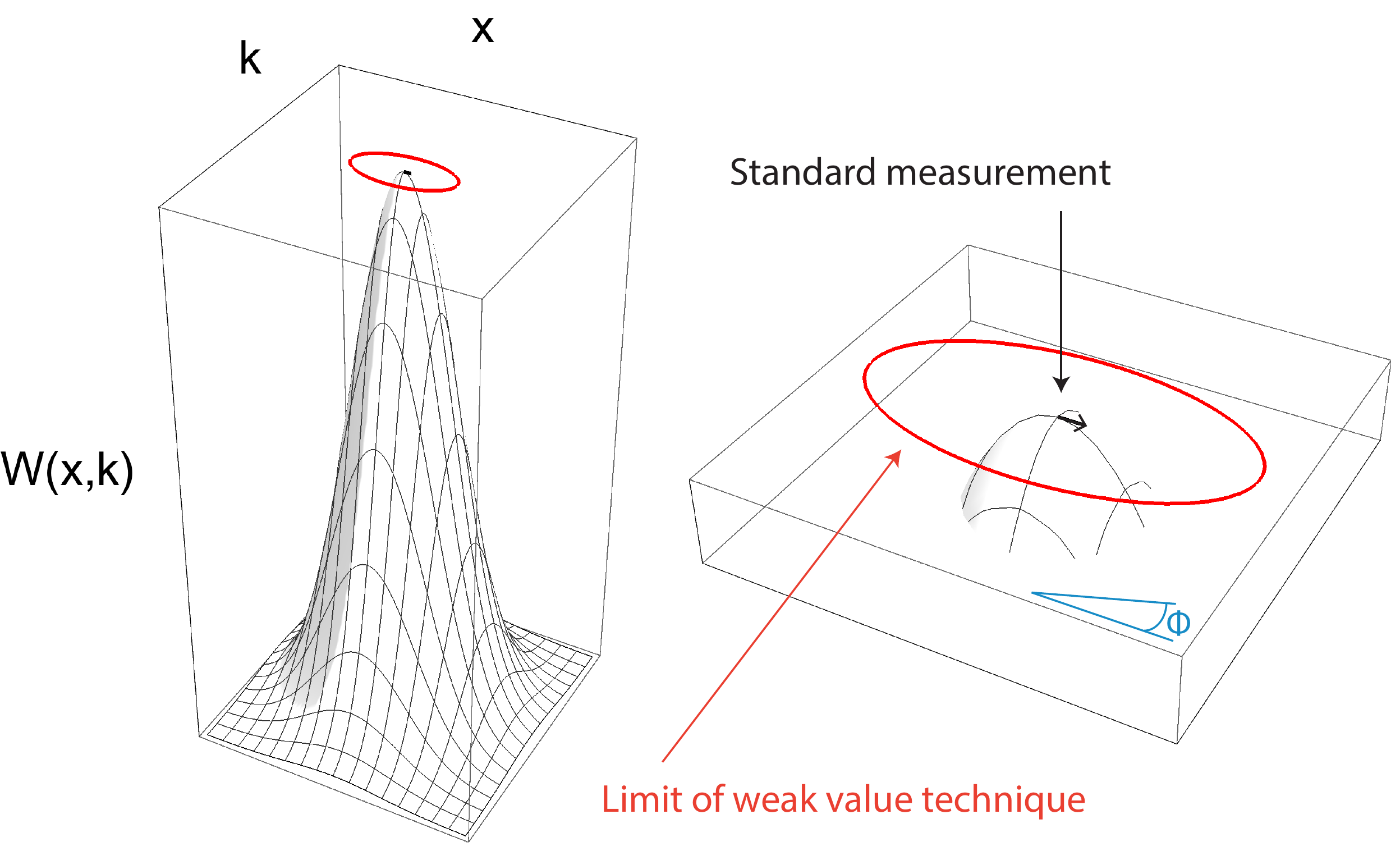}
\caption{\label{wigner}We represent the initial quantum state of the meter via a Wigner function. A standard measurement always succeeds and results in a small shift in the $x$ direction (black arrow, final state show in Figure 4).  The weak value technique is postselected, and allows for larger shifts in any direction $\phi$ in phase space. For the linear regime considered by AAV to hold, the weak value shift must lie well within the red curve, which is a parametric ellipse traced out by the maximum of the Wigner function when $g|A_w|=\sigma$ is substituted into Equation (\ref{wvshift}) and $\phi$ is allowed to vary. The characteristics of weak value experiments offer no fundamental advantage under perfect detection, but can alleviate certain types of detector imperfection. In this figure we take $\sigma=1,g=0.1,\lambda_*=1$.}
\end{figure}

We will consider a simple model of a qudit (the principal system) coupled to a single continuously varying degree of freedom (the meter), prepared in a real-valued Gaussian state:
\begin{eqnarray}
|i\rangle\otimes |m\rangle=|i\rangle \otimes \int dx \psi(x)|x\rangle =|i\rangle \otimes \int dk \phi(k)|k\rangle,  \nonumber\\
\psi(x)=\frac{1}{(2\pi)^\frac{1}{4}\sqrt{\sigma}}e^{-\frac{x^2}{4\sigma^2}},\nonumber \\
\phi(k)=\left(\frac{2}{\pi}\right)^{\frac{1}{4}}\sqrt{\sigma}e^{-k^2 \sigma^2}.
\label{gwf}
\end{eqnarray}
$\psi(x)$ represents the position wavefunction of the meter, while $\phi(k)$ describes the distribution over values for the conjugate momentum. $\sigma$ is a parameter controlling the uncertainty in $x$; and since Gaussian states saturate the Heisenberg uncertainty principle it also controls the uncertainty in $k$, but with an inverse relationship. To develop our visual depiction of weak-value experiments, we make use of the Wigner function description~\cite{Case2008} of $\psi$:
\begin{eqnarray}
W_m(x,k):=&\frac{1}{2\pi}\int \psi\left(x+\frac{y}{2}\right)\psi^*\left(x-\frac{y}{2}\right)e^{-iky}dy\nonumber\\
=&\frac{1}{2\pi}\int \phi\left(k+\frac{u}{2}\right)\phi^*\left(k-\frac{u}{2}\right)e^{ixu}du\nonumber\\
=& \frac{1}{\pi} e^{-2\sigma^2 k^2}e^{-\frac{x^2}{2\sigma^2}}.
\end{eqnarray}
The Wigner function may be thought of as a representation of the wave-function that allows one to simultaneously visualise both position ($x$) and momentum ($k$) variables. When $x$ is a component of the transverse position of a quantum particle (e.g. a photon or spin) propagating along $z$, $k$ is the corresponding component of the transverse wavevector (with the other transverse coordinate integrated out). Despite the possibility of negative regions (for a general Wigner function) precluding the interpretation $W(x,k)$ as a genuine probability distribution, it has the attractive property that its marginal distributions $\tilde{p}(k)=\int W(x,k)dx$ and $p(x)=\int W(x,k)dk$ correspond to those derived in the usual fashion, via the Born rule, for measurements of $k$ and $x$ respectively. They are therefore bona fide probability distributions, in the sense that they are positive and integrate to unity. Since the initial state of the meter is normalised, we thus naturally have that the volume of the initial Wigner function $\int\int W(x,k)dxdk=1$. See Figure \ref{wigner}.

\subsection{Weak value technique}
By applying equation~(\ref{wva}) to (\ref{gwf}), the Wigner function of the meter after the weak interaction and conditioned on successful postselection of the system into $|f\rangle$ will be:
\begin{eqnarray}
W_{A_w}(x,k)= |\langle f|i\rangle|^2W_m(x-g\mathfrak{Re}A_w,k- \frac{g}{2\sigma^2}\mathfrak{Im}A_w).
\label{wvshift}
\end{eqnarray}
This can be realised by spotting that the real and imaginary components of the weak value have actions $e^{-ig\mathfrak{Re}A_w\hat{k}}$ (unitary) and $e^{g\mathfrak{Im}A_w\hat{k}}$ (non-unitary) which mutually commute. This expression extends beyond the simple case of a real Gaussian function that we consider here but not to arbitrary complex valued functions: for a comprehensive treatment, see Ref~\cite{KofmanAshhabNori2012}. 

The approximation leading to Equation (4) requires $g|A_w| \ll \sigma$~\cite{DuckStevensonSudarshan1989}; a condition that is depicted as a red ellipse in Figure 1. The probabilistic nature of the protocol is reflected in $W_{A_w}$ being sub-normalised: the volume under the post-selected Wigner function is less than one and equal to the success probability. 

The Wigner function representation implies that a generally complex weak value will introduce a shift in an oblique direction in phase space. This raises the question of using `rotated' phase space observables to detect the shift. Define 
$$
\hat{s}_\theta : = \hat{x}\cos\theta + \hat{k}\sin\theta.
$$
The distribution for this rotated observable may be extracted from the Wigner function by integrating over its conjugate $\hat{t}_\theta : = -\hat{x}\sin\theta + \hat{k}\cos\theta$~\cite{BreitenbachSchillerMlynek1997}:
\begin{eqnarray}
p_\theta(s_\theta)&=\int_{-\infty}^\infty dt_\theta W_m(s_\theta\cos\theta-t_\theta\sin\theta,s_\theta\sin\theta+t_\theta\cos\theta)\nonumber\\
&=\frac{\sqrt{2}\sigma}{\sqrt{4\pi\sigma^4\cos^2\theta+\pi\sin^2\theta}}\exp\left\{-\frac{s_\theta^2}{2\sigma^2(2\sigma^2\cos^2\theta+\frac{1}{2\sigma^2}\sin^2\theta)}\right\}.
\end{eqnarray}
This is once more a Gaussian distribution, which can be thought of as a `view' of $W(x,k)$ from an oblique angle. The variance $\sigma_\theta^2$ is now a function of the $x$ and $k$ variances. One can show that after the weak value technique one is left with once again a simple shift:
\begin{eqnarray}
p_\theta(s_\theta)\rightarrow |\langle f|i\rangle|^2p_\theta(s_\theta - [g|A_w|\cos\phi\cos\theta+\frac{g}{2\sigma^2}|A_w|\sin\phi\sin\theta]).
\label{generalshift}
\end{eqnarray}
where we have written the weak value in polar form:
\begin{eqnarray}
\mathfrak{Re}A_w=|A_w|\cos\phi\nonumber\\
\mathfrak{Im}A_w=|A_w|\sin\phi.
\end{eqnarray}
\subsection{Standard measurement}
We will define a \emph{standard measurement} by setting $|i\rangle=|f\rangle=|i_*\rangle$ to the eigenstate of $A$ with eigenvalue of greatest magnitude. This scheme succeeds deterministically, and so results in a Wigner function of unit volume:
\begin{eqnarray}
W_{\textrm{std}}(x,k)= W_m(x- g\lambda_*,k)\nonumber\\
\Rightarrow p_{\textrm{std}}(x)=p(x-g\lambda_*).
\end{eqnarray}
Here $\lambda_* =\textrm{arg}\,\textrm{max}_\lambda\{|\lambda|:\hat{A}|i\rangle=\lambda|i\rangle\}$. The two strategies we consider can thus both be thought of as \emph{restricted} shifts of the initial Wigner function in phase space. Figure 1 shows the shift corresponding to the case where $\lambda_*=1$ with a black arrow. 

In depicting both standard and weak value techniques in the same plots (below in Figures 2 and 4 in particular), we have avoided dramatic examples such as $A_w=100$ for several reasons. Firstly, it is somewhat of a distraction that we wish to demote in favour of the more subtle but ultimately more useful properties of weak value techniques. Secondly, it is difficult to show such examples without adversely affecting our visual picture. Thirdly, and most importantly, our intuition is actually much better served by realising that the red ellipse in Figure~\ref{wigner}, which represents the very limit of validity of AAV's approach, is fixed by $\sigma$. Very large weak values are of course still possible, but only by ensuring $g$ is smaller and smaller. In this scenario, the black arrow shrinks while the red ellipse remains constant, meaning that the `amplification factor' $|A_w|/\lambda_*$ becomes very large, \emph{while the size of the shift itself must actually remain small}. This has previously been described as the requirement that the signal-to-noise ratio per data point must be low~\cite{Kedem2012}.
\section{Precision of estimating $g$}
Whenever a quantity of interest is not directly measured, it is necessary to process the raw data in order to `measure' -- or more strictly `estimate' -- that quantity. For instance, $g$ is not directly measured, but inferred from repeated measurements of $x$. The data are subsequently processed thus:
\begin{eqnarray}
\tilde{g}:=\frac{\langle x \rangle}{\lambda_*}.
\label{estimator}
\end{eqnarray}
Here angled brackets stand for the sample average. The rules governing the data processing are known as the `estimator': above we used a tilde to denote the estimator for $g$ in the standard technique. For the weak value technique, the procedure is similar, but $x$ is replaced by $s_\theta$ and $\lambda_*$ is replaced by the coefficient of $g$ in the argument of $p_\theta$ of Equation~(\ref{generalshift}). Because the data (the input to the estimator function, $x$) are random variables, the estimates (the output of the estimator function, $\tilde{g}$) are also random variables. We will assume that all estimators under consideration are unbiased, meaning that they give the correct answer on average\footnote{It is clear that our simple estimator is exactly unbiased for the standard strategy, but only approximately unbiased for the weak value technique. The latter fact is commonly overlooked.}, i.e. $\langle \tilde{g}\rangle=g$.

As alluded above, the classical Fisher information about the coupling constant $g$:
\begin{eqnarray}
F_g:=\int \frac{(\partial_g p_\theta(s_\theta))^2}{p_\theta(s_\theta)}ds_\theta
\end{eqnarray}
is the central figure of merit of classical parameter estimation. The Cram\'er Rao bound states that in the limit of many trials, the \emph{precision} (or \emph{uncertainty}, or \emph{standard deviation}) of an unbiased estimator for $g$ will be lower-bounded by the reciprocal of $\sqrt{F_g}$~\cite{Van-Trees1968}. The higher $F_g$, the better the precision, and the Cram\'er Rao bound can be saturated by \emph{efficient} estimation strategies: maximum likelihood, for example. Owing to the simplicity of the model considered here, the estimator given in Equation (\ref{estimator}) is indeed efficient for ideal detection, but requires modification when detector imperfections are included~\cite{KneeMunro2015b}.

The Fisher information measures the information content of a probability distribution $p_\theta(s_\theta)$, which can be derived from the Wigner function upon fixing a measurement. Under our formalism, any phase space quadrature may be considered, or indeed noisy implementations thereof. The choice of measurement thus influences $p_\theta(s_\theta)$ and thereby also influences $F_g$: so we may use the latter to evaluate the suitability of different measurement schemes.

Because we consider a single parameter, the maximum classical Fisher information about $g$ (when considering all possible measurements) of the joint-system meter state immediately after the interaction is given by the quantum Fisher information~\cite{BraunsteinCaves1994,Helstrom1969,Paris2009}:
\begin{eqnarray}
H_g  [|i\rangle\otimes|m\rangle]= \textrm{Var}(A\otimes \hat{k})=\langle \hat{A}^2\rangle\langle \hat{k}^2\rangle -\langle \hat{A}\rangle^2\langle \hat{k} \rangle ^2 =  \frac{\langle \hat{A}^2 \rangle}{\sigma^2}=\frac{\langle i |\hat{A}^2|i \rangle}{\sigma^2}.
\end{eqnarray}
$H_g$ is therefore a property of the initial quantum states alone, once the Hamiltonian has been fixed. It is clear that the smaller $\sigma$ -- the narrower the distribution in space -- the more information can be extracted. We will consider different measurement schemes that can harvest this maximal information in different ways, by channelling it into different parts of parameter space. That is, we will consider how close one can get to saturating
\begin{eqnarray}
F_g\leq H_g.
\end{eqnarray}
In general, if we consider system and meter to be a single indivisible quantum system, harvesting all of the information would require entangled preparation and entangled measurement. However, for the case where the average momentum of the meter is zero as it enters the weak interaction (i.e we take $\langle \hat{k} \rangle =0$ in accordance with Equation~(\ref{gwf})), factorable preparations and measurements will suffice.\footnote{A notable departure from this paradigm is when $\hat{k}$ is replaced with the particle number operator, which clearly will take nonzero values in general. This can unlock Heisenberg scaling in the Fisher information contained in the postselection probability $|\langle f | i \rangle|^2$ itself~\cite{ZhangDattaWalmsley2015,JordanTollaksenTroupe2015,ChenAharonSun2016}.} It follows from the convexity of the QFI~\cite{TothPetz2013}:
\begin{eqnarray}
H_g\left[\sum_j p_j |i_j\rangle\otimes|m_j\rangle\right] \leq \sum_j p_j H_g\Huge[|i_j\rangle\otimes|m_j\rangle\Huge]
\end{eqnarray}
that using mixed states cannot increase the Fisher information.

The chain rule of differentiation is a particularly effective tool in relating the Fisher information of a shifted distribution (which does not change shape) to properties of the shape of the distribution itself~\cite{KneeGauger2014}. Namely:
\begin{eqnarray}
F_g[p(s-\nu g)]=\nu^2F_s[p(s)] \nonumber\\
\Rightarrow F_g[p_\theta(s_\theta-\nu g)]= \frac{\nu^2}{\sigma_\theta^2}
\label{keyeq}
\end{eqnarray}
where the last step follows from the Gaussian shape of the marginal distribution.

\subsection{Optimal protocols}
It is straightforward to see that our \emph{standard measurement} sets a high Quantum Fisher information, and then extracts all of it. Recall that we set $|i\rangle=|f\rangle$ to the eigenstate of $A$ with highest magnitude eigenvalue $\lambda_*$, and then measure $x$:
\begin{eqnarray}
F_g[p_{std}(x)]&=\frac{\lambda_*^2}{\sigma^2}.
\end{eqnarray}
Full information is harvested, but one is `stuck' with measuring $x$ and with having a high flux onto the detector.
Our weak value technique, on the other hand, uses arbitrary initial and final states.
Rotated quadratures have 
\begin{eqnarray}
F_s[p_\theta(s_\theta)]=\frac{1}{\sigma_\theta^2}=\frac{1}{\sigma^2\cos^2\theta+\frac{1}{4\sigma^2}\sin^2\theta}.
\end{eqnarray}
We may use this formula, along with Equations~(\ref{wvshift}) and~(\ref{keyeq}) to evaluate a corrected Fisher information for a weak value shift for in an arbitrary direction $\phi$ and arbitrary measurement direction $\theta$ in phase space. To perform this calculation, we will normalize the distribution before calculating the Fisher Information and then correct it by the postselection probability. This is appropriate given the additivity of Fisher information: $N$ independent experiments enjoy a total Fisher information of $NF$, or in the case of weak value experiments a total of $|\langle f | i \rangle|^2NF$. We have
\begin{eqnarray}
|\langle f | i \rangle|^2F_g[p_{\textrm{wv}}(s_\theta)]
=|\langle f | i \rangle|^2F_g[p_\theta(s_\theta - [g\mathfrak{Re}A_w\cos\theta+\frac{g}{2\sigma^2}\mathfrak{Im}A_w\sin\theta])\nonumber\\
=|\langle f | i \rangle|^2|A_w|^2\left(\cos\theta\cos\phi+\frac{\sin\theta\sin\phi}{2\sigma^2}\right)^2F_s[p(s_\theta)].
\end{eqnarray}
We may now show that this information is always lower than than of the standard measurement:
\begin{eqnarray}
|\langle f | i \rangle|^2F_g[p_{\textrm{wv}}(s_\theta)]\leq \frac{|\langle f | i \rangle|^2 |A_w|^2}{\sigma^2}=\frac{|\langle f |A|i\rangle|^2}{\sigma^2}\label{ineq1}\\ \leq \frac{\langle \hat{A}^2\rangle}{\sigma^2}=H_g [|i\rangle\otimes|m\rangle]\label{ineq2}\\
\leq\frac{\lambda_*^2}{\sigma^2} =F_g[p_{\textrm{std}}(x)]=H_g [|i_*\rangle\otimes|m\rangle].
\label{ineq3}
\end{eqnarray}
The first inequality is proved in the appendix, along with a prescription for saturating it. The second inequality follows from the Cauchy-Schwarz inequality~\cite{KneeGauger2014}. The third inequality follows from the selection of the optimal initial state $|i_*\rangle$ by the standard measurement strategy. It is possible to get close to saturating these inequalities whilst maintaining a large weak value~\cite{AlvesEscherMatos-Filho2015,VizaMartinez-RinconAlves2015}. This means that almost all of the Fisher information may be concentrated into an unlikely or `dark' detector mode. We will not focus so much here on what combination of initial and final states, coupling parameters and meter states are necessary to approach equality in Equations~(\ref{ineq1}-\ref{ineq3}), save to note that if $\hat{A}^2=\mathbb{I}$ then $|f\rangle=\hat{A}|i\rangle$ will saturate the second condition and implies a real weak value. The question has been considered outside of the linear regime we consider~\cite{KneeGauger2014,Knee2014}. Instead the important point is that one may sacrifice only a small amount of information at this stage of the discussion: the sacrifice may well be considered negligible in comparison to other benefits that we describe below.

\section{Protecting information by alleviating technical noise}
So far we have considered `ideal' detection. This is to be understood as a projective and orthogonal measurement that one would typically find in a quantum mechanics text book: an Hermitian observable $\hat{x}$ or $\hat{k}$, for example. Such an idealised measurement offers unlimited precision when the number of repetitions $N$ tends to infinity. For finite trials, we get a finite precision; but this is a consequence of the quantum noise in the prepared state $|m\rangle$ (often referred to as `shot noise') rather than the detector. If the initial state was noise-free (which would correspond to the limit of a highly `squeezed' Wigner function with $x$ Dirac-delta distributed) an ideal detector would offer `perfect precision' -- an exact estimate with zero uncertainty after a single trial. Otherwise, non-ideal detectors are modelled by different kinds of coarse-grained approximations to ideal detectors, and imply a loss of information.  

We will consider the following examples of technical noise: (1) transverse jitter (2) pixelation and (3) saturation. Each will be defined by a transformation which relates the ideal distribution $p(s)$ to the imperfect one $p'(s)$. The results of this paper are to present the information obscuring effect of these transformations graphically, as Fisher information efficiency functions $\eta$, defined by: 
\begin{eqnarray}
F_g[p'(s)]=\eta F_g[p(s)].
\end{eqnarray}
These functions allow one to describe different imperfect measurements as $\eta\times100\%$  Fisher efficient, where $\eta=1$ recovers the ideal information. Let us consider the relative Fisher information for the weak value technique compared to the standard technique, both under noisy detection:
\begin{align}
\frac{|\langle f | i \rangle|^2N}{N}\frac{F_g[p'_{\textrm{wv}}(s)]}{F_g[p'_{\textrm{std}}(x)]}=\left(\frac{\eta^{\textrm{wv}}}{\eta^{\textrm{std}}}\right)\frac{|\langle f | i \rangle|^2F_g[p_{\textrm{wv}}(s)]}{F_g[p_{\textrm{std}}(x)]}.
\end{align}
At best $F_g[p_{\textrm{wv}}(s)]$ for the weak value technique may be comparable to that of the standard measurement: then the second factor of this equation (the relative Fisher information under ideal detection) will be close to unity.  If $\eta$ favours the weak value technique, however, the first factor (the relative Fisher information efficiency) can be greater than unity and there is scope for practical advantages to be had.

Our central idea is that one can guide the Fisher information around in phase space.  In particular one can usher it away from danger; from regions of phase space where the information would be lost or degraded. One can concentrate a large proportion (near one hundred percent) of the available information into a low number of events, or even into a conjugate variable for less-noisy detection. One can guide the detector distribution away from a faulty or noisy part of the detector.

\begin{figure}
\centering
\includegraphics[width=12cm]{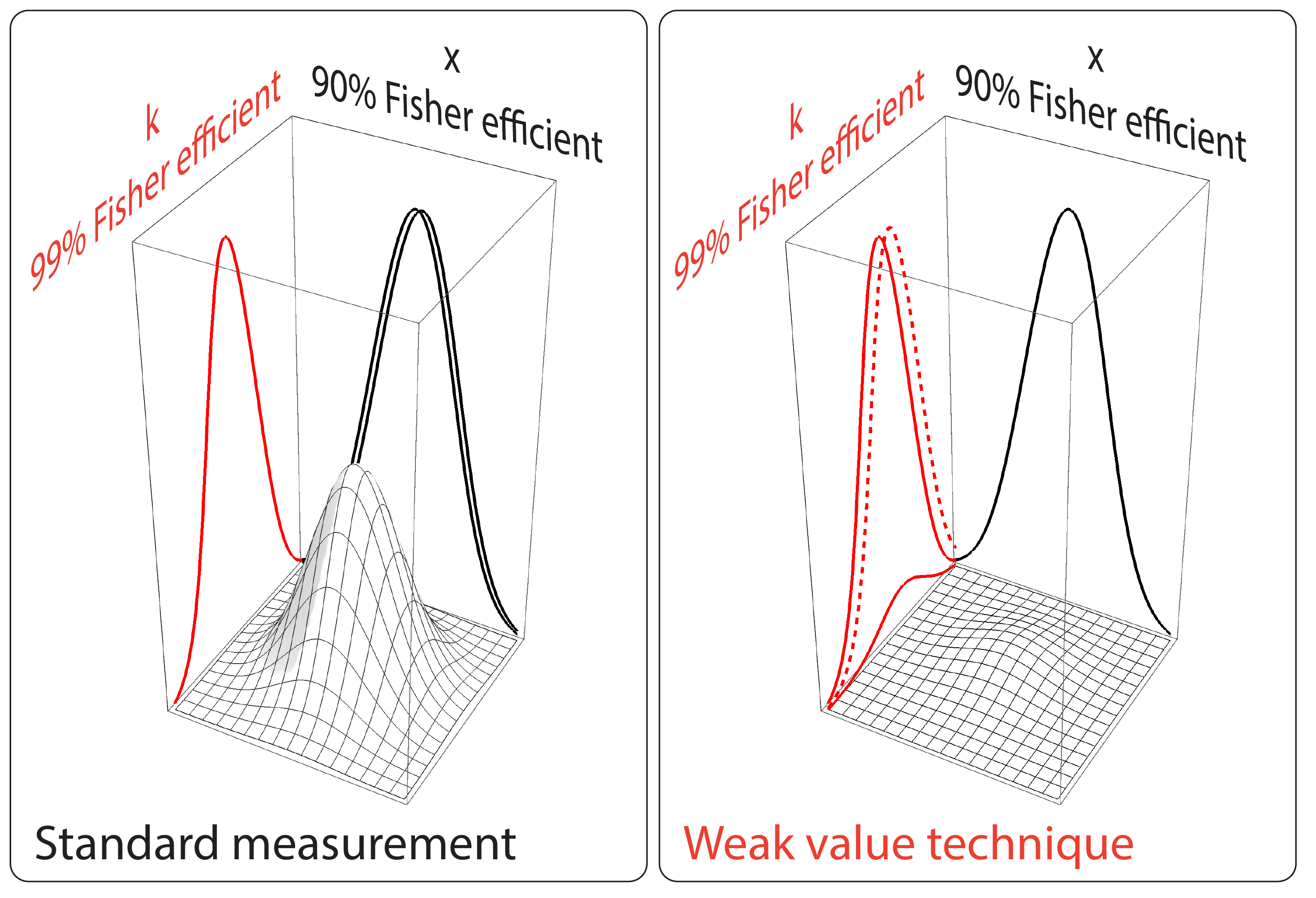}
\caption{\label{jitterfig}The Fisher information efficiency $\eta_{\textrm{jitter}}$ can differ for measurements in position ($x$, black) and momentum ($k$, red). Each panel shows the shifted Wigner distribution (wire mesh), along with the marginal distributions (solid black $p(x)$ and red $p(k)$) of both the initial and shifted meter states, drawn on the corresponding walls of the plot. Note how the standard technique is limited to changes in $x$ while $k$ remains unperturbed. In the weak value technique, $k$ carries Fisher information if pre and postselection are appropriately chosen (here $A_w=3i$ is pure imaginary). The Wigner function is attenuated due to the low success rate of postselection, and a renormalised marginal distribution is shown as a dashed line as a guide for the eye.  Transverse jitter reduces the Fisher information of a shifted Gaussian, but in a way that is independent of the magnitude of the shift and the intensity of the beam. It can be more or less severe in position or momentum space, however: in this example one could use a weak value technique to swap detection in $x$ (90\% Fisher efficient) for detection in $k$ (99\% Fisher efficient).}
\end{figure}
\subsection{Transverse jitter}
Consider that the detector is in a random motion, so that the actual distribution at the detector is 
\begin{eqnarray}
p'(s_\theta)=\int_{-\infty}^\infty p(s_\theta-y) \frac{e^{-\frac{1}{2}\frac{y^2}{2\zeta_\theta^2}}}{\sqrt{2\pi}\zeta} dy.
\end{eqnarray}
Under this model it is possible to show that~\cite{KneeGauger2014}:
\begin{eqnarray}
\eta_{\textrm{jitter}} : = \frac{F_g[p'(s_\theta)]}{F_g[p(s_\theta)]} = \left(1+\frac{\zeta_\theta^2}{\sigma_\theta^2}\right)^{-1}.
\end{eqnarray}
Note that there is no dependence on $s_\theta$, but only on $\sigma_\theta$ and $\zeta_\theta$. Since weak values may channel information into the $S_\theta$ variable (with $\theta$ not necessarily zero, which is impossible in standard measurements), this can be useful in the case where a more stable measurement of $s_\theta$ (than of $x$) is available~\cite{Kedem2012}. See Figure~(\ref{jitterfig}).
%%%
%%%
%%%
%%%
\subsection{Pixelation}
\begin{figure}
\centering
\includegraphics[width=14cm]{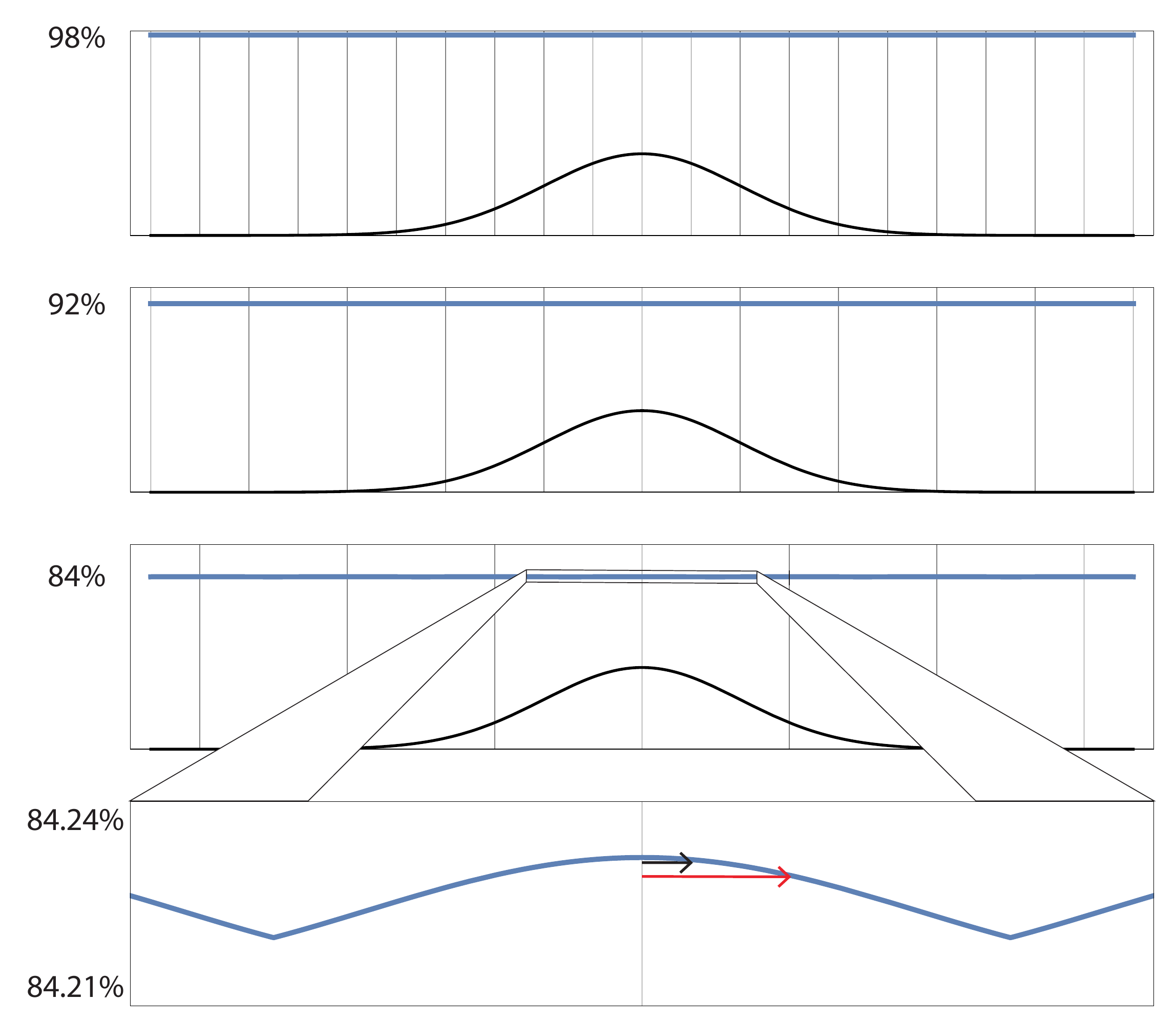}
\caption{\label{pixels}Plots of $\eta_{\textrm{pixel}}(\nu g)$ (blue) for pixel widths of $r=0.5,1,1.5$ respectively. This plot uses $\sigma=1$, with the continuous distribution before pixelation ($p(x)$, black) and the pixel boundaries (at $(n+\frac{1}{2})r$,  vertical grey bars) shown as a guide for the eye. The dependence on $s$ is weakly periodic and becomes weaker as the pixel size reduces. Even for ostensibly very large pixel sizes, the detection scheme is surprisingly effective: in these examples being at least 84\% Fisher efficient. For pixel sizes much smaller than $\sigma$, the detector is almost $100\%$ Fisher efficient. The bottom panel is a zoom of the $r=1.5$ example, and shows the shifts under a standard measurement (black) and weak value measurement with $A_w=3$ (red). Pixelation has mostly a negligible effect in general, but can even have a more detrimental effect on larger shifts (as in this example) depending on overall alignment of the detector~\cite{KneeGauger2014,KneeMunro2015b}. }
\label{pixel}
\end{figure}

Detection of continuous variables is often performed in a discretized fashion, such as with the pixel arrays found in modern cameras. We can model this by 
\begin{eqnarray}
\textrm{Pr}(n;\nu g)&=\int_{r(n)}^{r(n+1 )} p(s-\nu g) ds\nonumber \\
&=\frac{1}{2} \left(\text{erf}\left(\frac{\left(n+1\right) r-g \nu }{\sqrt{2} \sigma }\right)-\text{erf}\left(\frac{nr-g \nu }{\sqrt{2} \sigma }\right)\right),
\end{eqnarray}
where $r$ is the width of each pixel, labelled by integers $n$ and centred at $(n+\frac{1}{2})r$. In practice, this one-dimensional distribution is the result of summing over the $y$ direction of a two-dimensional pixel array. Here we have fixed the detector so that the pixel boundaries are aligned with the centroid of the initial meter state -- for the more general case of free alignment, see Ref.\cite{KneeGauger2014}. We can numerically compute
\begin{eqnarray}
\eta_{\textrm{pixel}}(\nu g)=\frac{F_g[\textrm{Pr}(n;\nu g)]}{F_g[p(s-\nu g)]};
\end{eqnarray}
a few examples are shown in Figure (\ref{pixel}). The efficiency is surprisingly high and only very weakly dependent on the parameters that distinguish standard measurements from weak value techniques: there is therefore likely that $\eta^{\textrm{wv}}_{\textrm{pixel}}\approx\eta^{\textrm{std}}_{\textrm{pixel}}\approx 1$. 

There are therefore two points to be made. Firstly, since $\eta_{\textrm{pixel}}$ is so robust to changing $r$, we conclude that pixelation does not represent much of a difficulty: there is limited scope for protecting information from this kind of imperfection. Secondly, $\eta_{\textrm{pixel}}$ does not depend very strongly on the size of the shift $\nu g$.  Recall that the distinguishing property of the weak value technique is the ability to organise for anomalously large shits: so that even if $\eta_{\textrm{pixel}}$ is low for the standard technique (say around 84\% as in Figure \ref{pixels}), it will be very similar for weak value technique. Therefore there is not much of an opportunity in this particular difficulty.

\subsection{Saturation}
\begin{figure}
\includegraphics[width=10cm]{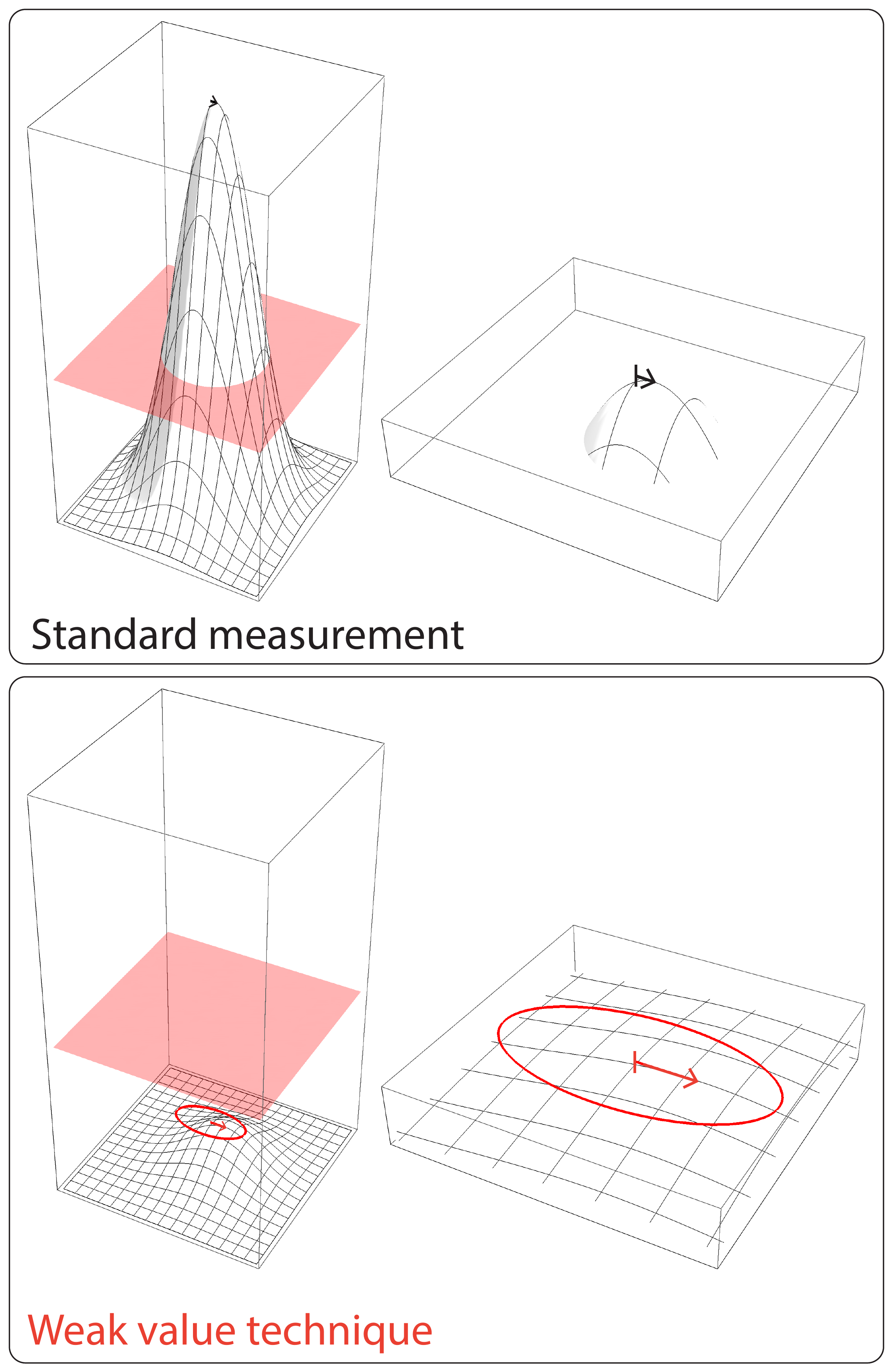}
\caption{\label{satfig}The problem of detector saturation is shown here in red as a maximum Wigner function volume (in our formalism this is equivalent to a maximum Wigner function height). When the Wigner function pokes through the red surface, this implies that the distribution at the detector will be `clipped', resulting in a loss of Fisher information. The weak value technique allows almost all of the available Fisher information (which is mostly concentrated into a distribution conditioned on an unlikely postselection event) to continue to a detector which would otherwise saturate. This figure uses $|\lambda_*|=1,A_w=3$. The black arrow and the red ellipse have the same meaning as in Figure 1.}
\end{figure}
Avoiding detector saturation is perhaps the most promising application of weak values in metrology. An information-theoretical advantage has been conjectured several times~\cite{HostenKwiat2008,Kedem2012,KneeGauger2014,Vaidman2014}: and a recent study by Harris, Boyd and Lundeen~\cite{HarrisBoydLundeen2017} gives these conjectures a theoretical underpinning. We refer the reader to~\cite{HarrisBoydLundeen2017} for a full analysis, but the rough argument is as follows: for a deterministic strategy, there will be some maximum photon flux for which the information will be degraded due to `bleaching' or saturation of the photodetectors. On the other hand, if $|\langle f | i \rangle|^2$ is small enough, then saturation will not be an issue and $\eta_{\textrm{sat}}\approx1$. 

In practice, it is likely that saturation problems will be avoided altogether (by reducing the intensity of the laser source, for example with a neutral density filter). It will therefore serve our purposes adequately to define 
\begin{align}
\eta_{\textrm{sat}}=\Theta(P_{\textrm{sat}}-|\langle f | i \rangle|^2)
\end{align}\
for $\Theta$ the Heaviside step function defined by 
\begin{align}
\Theta(z) = 
\begin{cases} 
      0 & z< 0 \\
      1 & z\geq 0. 
   \end{cases}
\end{align}
Of course in reality $\eta_{\textrm{sat}}$ may not fall to zero and may or may not feature a discontinuity at $P_{\textrm{sat}}$, but will monotonically decrease with $|\langle f | i \rangle|^2$ for a high enough input photon flux. We show our rough model pictorially in Figure~\ref{satfig}\footnote{Harris, Boyd and Lundeen report the surprising claim that saturation alone is not sufficient for an advantage with weak value techniques~\cite{HarrisBoydLundeen2017}. In their model, pixelation or generic detector noise are required, too. This is a consequence of their `soft' model of saturation which does not feature a discontinuity and maintains a one-one relationship between incident photon number and output photoelectron number in each pixel of the detector}.

\section{Conclusion}
By using the Wigner representation of the meter to describe weak value experiments, we have provided a simple picture of the flexibility of the method and how it compares to standard measurements. Working with real Gaussian meter states and in the linear regime considered by AAV, we have extended previous arguments~\cite{KneeGauger2014} concerning the Fisher information about the coupling parameter between system and meter: allowing for arbitrary complex weak values and measurements at oblique angles in phase space. Under ideal detection, the Fisher information  of the weak value technique will never exceed the Fisher information of the standard strategy without post-selection. We then considered the impact of three types of technical noise, namely (1) transverse jitter, (2) pixelation and (3) photo-saturation of the detector.  The Fisher information efficiency that characterises these imperfections may in some cases be higher for the weak value technique than for the standard strategy, in which case weak value techniques would give improved precision. Whilst this might conceivably be the case for (1) and (2), (3) is the killer application of weak value techniques. By concentrating almost all of the available quantum Fisher information into a low number of events, it can amplify the parameter to be measured when detector saturation would otherwise be a limitation. As long as the cost of generating a photon is considered lower than the cost of detecting one, weak values therefore offer greater precision per unit cost than standard measurements. This conclusion invites experimental exploration and implementation. 
\section{Appendix}
Let $F=F_g[p_\theta(s_\theta - [g\mathfrak{Re}A_w\cos\theta+\frac{g}{2\sigma^2}\mathfrak{Im}A_w\sin\theta])$. Assume for contradiction that $\frac{\sigma^2 F}{|A_w|^2}>1$. By algebraic manipulations, this condition is equivalent to 
\begin{eqnarray}
\left(2\sigma \tan\phi - \frac{1}{\sigma}\tan\theta \right)^2< 0 \Rightarrow\perp.
\end{eqnarray}
hence we must conclude that $F\leq |A_w|^2/\sigma^2$. Furthermore, the optimum measurement to saturate this inequality  is simply deduced:
\begin{eqnarray}
\theta\rightarrow \arctan(2\sigma^2\tan\phi),
\end{eqnarray}

\begin{acknowledgements}
GCK was supported by the Royal Commission for the Exhibition of 1851. We thank Natalia Ares for helpful comments on this manuscript.
\end{acknowledgements}

\bibliographystyle{unsrt}
\bibliography{/Users/georgeknee/Documents/paper_library/gck_full_bibliography}  % name your BibTeX data base

%\begin{thebibliography}{}
%%
%% and use \bibitem to create references. Consult the Instructions
%% for authors for reference list style.
%\end{thebibliography}

\end{document}